\documentclass[jcp,aip,10pt,floatfix,twocolumn,showpacs]{revtex4-1} 
\usepackage{color,soul}
\usepackage{graphicx}
\usepackage{subcaption}
\usepackage{siunitx}
\usepackage[breaklinks=true,colorlinks=true,linkcolor=blue,urlcolor=blue,citecolor=blue]{hyperref}

\usepackage{amsmath, amsthm, amssymb}
\usepackage{amsthm}
\usepackage{microtype}
\usepackage{caption}
\usepackage{multirow}
\usepackage{float}
\usepackage[normalem]{ulem}
\usepackage{soul}
\usepackage{makecell}
\usepackage[table]{xcolor}
\usepackage{ragged2e}
\definecolor{Gray}{gray}{0.85}
\definecolor{LightCyan}{rgb}{0.88, 1, 1}
\definecolor{Apricot}{rgb}{0.98, 0.81, 0.69}
\usepackage{threeparttable}
\usepackage{siunitx}
\newcommand{\be}{\begin{equation}}
\newcommand{\ee}{\end{equation}}
\newcommand{\bea}{\begin{eqnarray}}
\newcommand{\eea}{\end{eqnarray}}

\usepackage[rightcaption,]{sidecap}
\sidecaptionvpos{figure}{c}
\usepackage[utf8]{inputenc}
\usepackage{amsmath}
\usepackage{amssymb}
\usepackage{graphicx}
\usepackage{hyperref}
\usepackage[export]{adjustbox}
\usepackage[update]{epstopdf}
\usepackage{multirow}
\usepackage{placeins}
\usepackage{xr}
\begin{document}

\title{Density-driven reentrant polymer transitions via saturable bridging crowders}
\author{Monmee Phukan}
\email{monmee@imsc.res.in}
\affiliation{The Institute of Mathematical Sciences, C.I.T. Campus,
Taramani, Chennai 600113, India}
\affiliation{Homi Bhabha National Institute, Training School Complex, Anushakti Nagar, Mumbai, 400094, India}

\author{Hitesh Garg}
\email{hiteshgarg@imsc.res.in}
\affiliation{The Institute of Mathematical Sciences, C.I.T. Campus,
Taramani, Chennai 600113, India}
\affiliation{Homi Bhabha National Institute, Training School Complex, Anushakti Nagar, Mumbai, 400094, India}

\author{Satyavani Vemparala}
\email{vani@imsc.res.in}
\affiliation{The Institute of Mathematical Sciences, C.I.T. Campus,
Taramani, Chennai 600113, India}
\affiliation{Homi Bhabha National Institute, Training School Complex, Anushakti Nagar, Mumbai, 400094, India}

\date{\today}
\begin{abstract}
Reentrant coil-globule-coil transitions, in which a polymer collapses and then reexpands as a single parameter is varied, have been observed across diverse soft matter systems, yet the minimal ingredients required to produce them remain unclear. Using molecular dynamics simulations of coarse-grained polymers interacting with a single species of attractive crowder, we show that crowder volume fraction $\phi_c$ alone is sufficient to drive a complete reentrant transition. At low $\phi_c$, crowders bridge distant monomers and drive cooperative collapse; at high $\phi_c$, saturation of monomer binding sites suppresses bridging connectivity and produces reentrant expansion. This density-driven transition is absent with purely repulsive crowders, which produce only monotonic compaction while preserving self-avoiding walk (SAW) chain statistics. In contrast, bridging breaks SAW universality: the rescaled size distributions no longer collapse onto a universal curve, and the conformational distributions trace the full coil-globule-coil trajectory as $\phi_c$ is varied. For charged polymers with explicit counterions, electrostatics amplifies rather than suppresses reentrance: bridging crowders displace counterions from the chain, and upon saturation the unscreened backbone charges drive expansion well beyond the original chain size. Saturable geometric bridging thus emerges as a minimal mechanism linking reentrant phenomena across neutral and charged polymers in crowded environments.
\end{abstract}

\maketitle

\section{Introduction}

Reentrant transitions, in which a system collapses at intermediate conditions but reexpands at higher concentrations of the same parameter, appear regularly across soft matter and biology~\cite{milin2018reentrant,banerjee2017reentrant,feng2015re,truzzolillo2018overcharging,marzi2015depletion}. In polymer physics, the canonical manifestation is the coil-globule-coil transition~\cite{RubinsteinColby2003,GroKho1994}, in which a single chain compacts and then reswells as a control parameter is varied monotonically. The microscopic origins of reentrance vary widely across systems, involving competitive solvation, electrostatic correlations, multivalent binding, or depletion interactions, yet a common motif recurs: the same interactions that drive collapse become less effective as concentration increases~\cite{mukherji2017depleted,Mukherji2014NatComm}.

Neutral polymers offer some of the clearest early examples. Poly(\emph{N}-isopropylacrylamide) (PNIPAM) dissolves readily in pure water or pure methanol, yet collapses in their mixtures at intermediate methanol fractions before reswelling at high methanol content~\cite{Zhang2001PRL}; the mechanism is preferential adsorption of cosolvent molecules that bridge chain segments enthalpically, with an efficiency that is finite and non-monotonic in composition~\cite{Mukherji2014NatComm,Scherzinger2022SoftMatter}. Related transitions occur in hydrogels and polymer networks where hydrophobic and hydrogen-bonding interactions compete~\cite{Okay2021Gels,oh2013molecular}. In electrically charged systems, DNA and synthetic polyanions collapse in moderate concentrations of multivalent counterions through ionic bridging and charge neutralization, but redisperse at higher ion concentrations via overcharging or charge inversion~\cite{Raspaud1998BJ,Olvera1995JCP,Zhang2008PRL}, with ion correlations and underscreening identified as additional mechanisms that can restore long-range repulsion at high ionic strength~\cite{Liu2020JCIS,Robertson2023PCCP}. Proteins show the same pattern: lysozyme aggregates near intermediate multivalent ion concentrations but redisperses at higher levels~\cite{Zhang2008PRL}. In colloid-polymer mixtures, non-adsorbing polymers induce depletion attractions that stabilize colloidal crystals at intermediate concentrations but produce reentrant melting at higher polymer loading~\cite{Meng2017NPJ,lekkerkerker2011depletion}, while adsorbing polymers cause bridging flocculation at intermediate coverage that gives way to steric stabilization once particle surfaces are saturated~\cite{ScheutjensFleer1979,Tripathy2017AdvColloid}. Recent experiments on intrinsically disordered proteins in crowded environments have similarly revealed non-monotonic conformational responses that depend on the sign and range of protein-crowder interactions~\cite{Balu2022SciAdv,Holehouse2024NatRevMolCellBiol}. Across all of these systems, collapse is strongest when attractive or bridging interactions are most effective, and reexpansion follows when those interactions are suppressed, saturated, or reversed. This suggests that reentrance may emerge generically whenever the efficiency of multivalent interactions varies non-monotonically with concentration, without requiring complex chemistry or long-range electrostatics.

To make this intuition precise requires examining how crowding affects polymer conformations at the microscopic level. Crowding effects arise from two distinct contributions: steric volume exclusion (purely repulsive and entropic in origin) and non-specific soft attractive (enthalpic in origin) interactions~\cite{grassmann2024computational,alfano2024molecular,rivas2016macromolecular,zhou2008macromolecular,ellis2003join,zimmerman1993macromolecular}. In the purely repulsive limit, Kang \emph{et al.}~\cite{Kang2015PRL} established that crowders compact a polymer through osmotic pressure, with compaction controlled by the dimensionless ratio $x = \bar{R}_g(0)/D$, where $\bar{R}_g(0)$ is average radius of gyration of the polymer without crowders and $D$ is the inter-crowder spacing; crucially, this depletion-driven response is strictly monotonic, and the rescaled size distribution $P(t)$ collapses onto a universal self-avoiding walk (SAW) curve~\cite{DeGennes1979,Lhuillier1988JPhys} at all densities. Attractive polymer-crowder interactions change this picture: Huang and Cheng~\cite{HuangCheng2021JPolSci} reported non-monotonic conformational transitions as crowder size is varied at fixed interaction strength; conformational phase diagrams have been mapped as functions of monomer-crowder attraction and intrapolymer interaction for both neutral and charged chains~\cite{Garg2023JCP,Tripathi2023JCP}, identifying extended, bridged-collapsed, and counterion-collapsed phases; and crowder-size dependence has been shown to generate complex collapse-extension-recollapse sequences even for neutral chains~\cite{garg2025polymer,Nayar2020PCCP,Nayar2023JPCB}. In a related context, a recent theoretical treatment of polyelectrolyte reentrant condensation identified a ``gluonic'' bridging mechanism, in which shared multivalent ions form physical crosslinks between ionic monomers, as the essential driver of both collapse and re-entry~\cite{Yong2024Biomacromolecules}. In all of these studies, however, the control parameters are interaction strength, crowder size, or cohesion at fixed crowder density; the reentrant response driven by crowder volume fraction $\phi_c$ alone has not been characterized.

The central question of interest is whether crowder volume fraction alone, at fixed interaction strength, is sufficient to drive a complete reentrant coil-globule-coil transition, and if so, what minimal microscopic ingredients are required. Bridging interactions of the kind described above are not specific to any one chemistry: they arise wherever crowders simultaneously adsorb onto multiple sites of a larger macromolecule, a situation common in biology (cosolvent bridging, ion-mediated condensation, protein-DNA crosslinking) and in synthetic soft matter. If density-driven reentrance can be produced by a minimal model with a single crowder species at fixed interaction strength, it would indicate that the reentrant phenomena observed across these chemically diverse systems share a common geometric origin rather than requiring system-specific explanations. Such a model also permits a direct comparison with the depletion framework of Kang \emph{et al.}~\cite{Kang2015PRL}: does bridging preserve the SAW universality of chain statistics that depletion maintains, or does it drive the chain into new statistical regimes? And does the mechanism survive in charged polymers, neutralized by oppositely charged counterions, where electrostatic repulsion competes with short-ranged bridging attraction?~\cite{Manning1969JCP,srivastava2016polyelectrolyte}

Here we address these questions using molecular dynamics simulations of coarse-grained neutral and charged polymers interacting with a single species of neutral crowder, with crowder volume fraction $\phi_c$ as the sole control parameter. We characterize the density-driven conformational response across the full range of $\phi_c$, identify the microscopic mechanism underlying both collapse and reexpansion, and examine how backbone electrostatics modulates the transition. Our results establish saturable geometric bridging as a minimal and generic route to polymer reentrance and provide a unified framework linking neutral and charged systems.

\section{Model and Methods}\label{Sec-2}

We employ a coarse-grained bead-spring model for a flexible linear polymer of $N_m = 50$ identical monomer beads of diameter $\sigma_m$ and mass $m$, confined in a cubic simulation box of volume $V = L^3$ with $L = 30\sigma_m$. For neutral chains each monomer carries zero charge; for charged chains each monomer carries charge $-qe$. Monomer beads are connected by harmonic springs with bonded potential
\begin{equation}
    U_{b}(r) = \frac{1}{2}k(r - r_0)^2,
\end{equation}
where $k = 500\,k_BT\sigma_m^{-2}$ is the spring constant, $r$ is the distance between bonded beads, $r_0 = 2^{1/6}\sigma_m$ is the equilibrium bond length, $k_B$ is Boltzmann's constant, and $T$ is the temperature.

For charged polymers, electroneutrality is maintained by introducing $N_{ci} = N_m / Z$ explicit counterions, each of diameter $\sigma_m$, mass $m$, and charge $+e$, where the valency $Z = 1, 3$ for monovalent and trivalent counterions respectively. Non-bonded interactions (monomer--monomer, monomer--crowder, 
crowder--crowder) are described by the truncated and shifted 
Lennard-Jones (LJ) potential:
\begin{equation}
    U_{ij}(r) = 4\epsilon_{ij}\left[\left(\frac{\sigma_{ij}}{r}
    \right)^{12} - \left(\frac{\sigma_{ij}}{r}\right)^6\right] 
    - U_{ij}(r_c^{ij}), \quad r < r_c^{ij},
\end{equation}
where $i,j \in \{m,c\}$ denote monomer or crowder species. The parameters $\epsilon_{ij}$, $\sigma_{ij}$, and cutoff $r_c^{ij}$ are pair-dependent and listed in Table~\ref{tbl:LJ_parameters} for purely repulsive (depletion) and bridging interactions. All length scales are expressed in units of $\sigma_m$. In the bridging simulations reported here, the monomer-crowder attraction strength is fixed at $\epsilon_{mc} = 4\,k_BT$.

Neutral crowders $N_c$ of diameter $\sigma_c$ and volume fraction
\begin{equation*}
    \phi_c = \frac{N_c}{V}\frac{4}{3}\pi\left(\frac{\sigma_c}{2}\right)^3
\end{equation*}
are introduced into the simulation box. Values of $N_c$ for different parameter combinations studied are listed in Table~\ref{tbl:crowder_number}.

For charged systems, long-range electrostatic interactions are described by the Coulomb potential
\begin{equation}
    U_C(r) = \frac{q_1 q_2 e^2}{4\pi\varepsilon\varepsilon_0 r},
\end{equation}
where $\varepsilon_0$ is the permittivity of free space and $\varepsilon$ is the uniform relative dielectric constant of the medium. The charge density along the polymer chain is characterized by the dimensionless 
Manning parameter
\begin{equation}
    A = \frac{q^2 l_B}{r_0},
\end{equation}
where $l_B = e^2/(4\pi\varepsilon\varepsilon_0 k_BT)$ is the Bjerrum length. We set $A/A_c = 0.3$, a value at which the polymer is in an extended state; previous work has shown that charged chains at this coupling undergo collapse in the presence of attractive crowders~\cite{Tripathi2023JCP}.

The degree of crowding-induced compaction is characterized by the 
dimensionless parameter introduced by Kang \emph{et al.}~\cite{Kang2015PRL},
\begin{equation}
    x = \frac{\bar{R}_g(0)}{D},
\end{equation}
where $D$ is the mean inter-crowder spacing. Since the volume per 
crowder is $D^3 = (4\pi\sigma_c^3/3)/\phi_c$, one obtains 
$D = (4\pi/3)^{1/3}\sigma_c\,\phi_c^{-1/3}$, and hence
\begin{equation}
    x = \left(\frac{3}{4\pi}\right)^{1/3}\lambda\,\phi_c^{1/3},
\end{equation}
where $\lambda = \bar{R}_g(0)/\sigma_c$. Values of $x$ for the 
$(\lambda,\phi_c)$ combinations studied are listed in Table~\ref{tbl:dimless_parameter_x}. 
In the depletion limit, collapse requires $x \gg 1$; in the bridging 
regime studied here, reentrance occurs across the full range of 
$x$ values.

All simulations are performed using LAMMPS~\cite{LAMMPS}. Each system is first equilibrated for $10^6$ steps in the NPT ensemble to attain the target density, then run for $10^7$ steps in the NVT ensemble with periodic boundary conditions. Temperature is maintained at $T = 1.0$ via a Nos\'{e}-Hoover thermostat; equations of motion are integrated with the velocity-Verlet algorithm at timestep $\Delta t = 0.001\,\tau$, where $\tau = \sqrt{m\sigma_m^2/\epsilon}$. Electrostatic interactions are evaluated using the particle-particle particle--mesh (PPPM) method with a relative accuracy of $10^{-4}$. Ensemble averages and standard errors are obtained by block averaging. Selected simulations for extreme values of crowder densities $\phi=0.01,0.4$ were repeated using different random seeds for initial velocity. The observed trends remained consistent, indicating that the results are robust to the choice of seed.

\begin{table}[h]
\centering	
\caption{\label{tbl:LJ_parameters} Parameters for the LJ potential [see Eq.~(2)] between different pairs of particles. Here $m$ and $c$ represent monomers and crowders, respectively. }
\begin{ruledtabular}
\begin{tabular}{l c c c c}
 & \multicolumn{2}{c}{Depletion} & \multicolumn{2}{c}{Bridging} \\
\cline{2-3} \cline{4-5}
Pair & $\epsilon_{jk}$ & $r_c^{jk}$ & $\epsilon_{jk}$ & $r_c^{jk}$ \\
\hline
$m-m$ & 1.0 & $2^{1/6}$ & 1.0 & $2^{1/6}$ \\
$m-c$ & 1.0 & $2^{1/6}$ & $4.0$ & $2.5\,\sigma_{mc}$ \\
$c-c$ & 1.0 & $2^{1/6}$ & 1.0 & 2.5 \\
\end{tabular}
\end{ruledtabular}
\end{table}

\begin{table}[h]
\centering
	\caption{\label{tbl:crowder_number} Number of crowders ($N_c$) in the different simulations.}
	\begin{ruledtabular}
		\begin{tabular}{| c | c c c c |}
			\hline
			$\lambda$ & 6.9 & 3.6 & 1.8 & 0.9 \\
			\hline
			$\phi_c$ & \multicolumn{4}{ c |}{$N_c$}\\
			\hline
			0.1 & 10866 & 1528 & 191 & 24 \\
			0.2 & 21733 & 3056 & 382 & 48 \\
			0.3 & 32599 & 4584 & 573 & 72 \\
			0.4 & 43465 & 6112 & 764 & 95 \\
			\hline
		\end{tabular}
	\end{ruledtabular}
\end{table}

\begin{table}[h]
\centering
    \caption{\label{tbl:dimless_parameter_x} Variation of dimensionless parameter $x$ for a neutral polymer as a function of crowder size $\lambda$ and crowder density $\phi_c$.}
    \begin{ruledtabular}
        \begin{tabular}{| c | c c c c |}
            \hline
            $\lambda$ & 6.9 & 3.6 & 1.8 & 0.9 \\
            \hline
            $\phi_c$ & \multicolumn{4}{ c |}{$x=(3/4\pi)^{1/3}\lambda\phi_c^{1/3}$}\\
            \hline
            0.01 & 0.92 & 0.48 & 0.24 & 0.12 \\ 
            0.05 & 1.58 & 0.82 & 0.41 & 0.21 \\
            0.1 & 1.99 & 1.04 & 0.52 & 0.26 \\
            0.2 & 2.5 & 1.31 & 0.65 & 0.33 \\
            0.3 & 2.87 & 1.5 & 0.75 & 0.37 \\
            0.4 & 3.15 & 1.65 & 0.82 & 0.41 \\
        \end{tabular}
    \end{ruledtabular}
\end{table}

\section{Results}\label{Sec-3}
\subsection{Bridging-induced reentrance in neutral polymers}
\begin{figure}
\includegraphics[width=0.9\columnwidth]{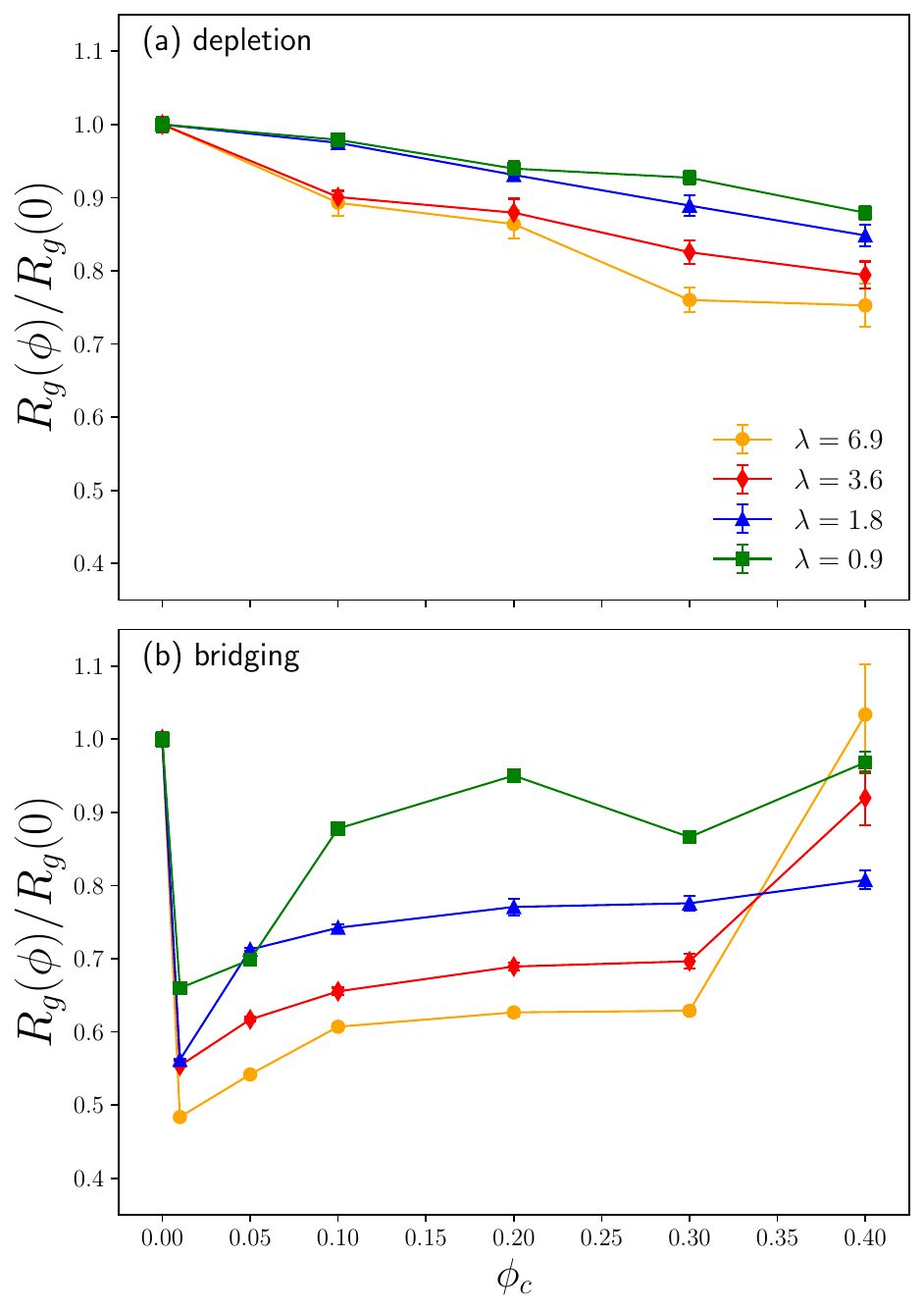}
\caption{Normalized radius of gyration $R_g(\phi_c)/R_g(0)$ of a neutral polymer as a function of crowder volume fraction $\phi_c$ for four size ratios $\lambda = R_g(0)/\sigma_c$ for (a) depletion regime (all interactions repulsive): $R_g(\phi_c)$ decreases monotonically with $\phi_c$ for all $\lambda$, with no reentrant behavior. (b) bridging regime (attractive monomer-crowder interactions, $\epsilon_{mc} = 4\,k_BT$): the polymer collapses at low $\phi_c$ and reexpands at high $\phi_c$ for all $\lambda$. Error bars represent standard errors from block averaging.}
\label{rg-wca-bridge}
\end{figure}

We first examined the conformational response of a neutral polymer chain to neutral crowders under two limiting interaction scenarios: purely repulsive interactions (depletion) and attractive polymer-crowder interactions (bridging). Figure~\ref{rg-wca-bridge} shows the normalized radius of gyration, $R_g(\phi_c)/R_g(0)$, as a function of crowder volume fraction $\phi_c$ for different values of the crowder size ratio $\lambda = R_g(0)/\sigma_c$, where $\sigma_c$ is the crowder diameter. The two panels contrast the conformational responses produced by repulsive and attractive crowder-monomer interactions.

In the depletion case [Fig.~\ref{rg-wca-bridge}(a)], all interactions between and among monomers and crowders are purely repulsive. $R_g(\phi_c)$ decreases monotonically with $\phi_c$ for all size ratios, consistent with the entropic compression of the chain by osmotic pressure from excluded crowders~\cite{Kang2015PRL}. The compaction is most pronounced for the smallest crowders (largest $\lambda$): at $\phi_c = 0.4$, the chain contracts to approximately $75\%$ of its free-solution size for $\lambda = 6.9$, compared to only $\sim\!88\%$ for $\lambda = 0.9$. This systematic $\lambda$-dependence is consistent with the scaling argument of Kang \emph{et al.}~\cite{Kang2015PRL}, wherein compaction is controlled by the ratio $x = \bar{R}_g(0)/D$ of polymer size to inter-crowder spacing $D$: smaller crowders at fixed $\phi_c$ yield a smaller $D$, hence larger $x$, and stronger osmotic compression.  For the depletion case, the polymer chain does not show any indication of reentrant behavior, for the parameter space considered.

The situation changes when attractive polymer-crowder interactions are introduced [Fig.~\ref{rg-wca-bridge}(b)]. At $\phi_c \approx 0.01$, the chain contracts to $\sim\!50\%$ of $R_g(0)$ for $\lambda = 6.9$ and to $\sim\!66\%$ for $\lambda = 0.9$, a much deeper and more abrupt collapse than seen at any density in the depletion regime. This rapid compaction at low crowder density reflects the strong adsorption affinity: when crowders are fewer and monomer binding sites are plentiful, each crowder can simultaneously contact multiple distant monomers, acting as an effective multivalent bridge. This collapse, in which the first few adsorbed crowders generate intrachain crosslinks that promote further compaction, is consistent with a cooperative transition (examined quantitatively through the conformational distributions in later section~\ref{subsection:chainstatneutral}).

As $\phi_c$ increases beyond this initial collapse, the chain remains in a compact bridged globule over a broad range of intermediate densities. At higher $\phi_c$, however, the polymer reexpands, with $R_g(\phi_c)/R_g(0)$ increasing for all $\lambda$ values. For $\lambda = 0.9$ and $\lambda = 3.6$, $R_g$ recovers to approximately its free-solution value by $\phi_c = 0.4$. For $\lambda = 6.9$, the chain reaches $R_g(\phi_c)/R_g(0) \approx 1.0$ at $\phi_c = 0.4$, indicating essentially complete recovery of the unperturbed chain size. The non-monotonic coil-globule-coil trajectory is thus present for all crowder sizes studied.

The reentrant transition depends on $\lambda$: small crowders (large $\lambda$) produce the deepest initial collapse and the most complete recovery, because their small size allows them to bridge across multiple monomers simultaneously at low density, but also to pack densely around individual monomers at high density, saturating bridging sites efficiently. Large crowders (small $\lambda$) produce shallower collapse and less complete recovery: their geometry limits both the efficiency of multivalent bridging at low $\phi_c$ and the degree of site saturation at high $\phi_c$. Across all $\lambda$ values, however, the non-monotonic response is present, establishing that the reentrant mechanism is robust to crowder size. These results strongly suggest two aspects. First, purely entropic depletion effects produce only monotonic, decelerating compaction and cannot account for reentrant behavior, consistent with the depletion framework of Kang \emph{et al.}~\cite{Kang2015PRL}. Second, attractive bridging interactions yield a non-monotonic conformational response: collapse at low crowder density driven by multivalent bridging, followed by a broad plateau of compact globule conformations, and reentrant expansion at high density as bridging valency is saturated. This full coil-globule-coil sequence emerges from a single homopolymer interacting with a single crowder species at fixed interaction strength, with $\phi_c$ as the sole control parameter.

\subsection{Monomer--crowder correlations reveal bridging and saturation mechanisms}
\begin{figure}
\centering
\includegraphics[width=0.95\columnwidth]{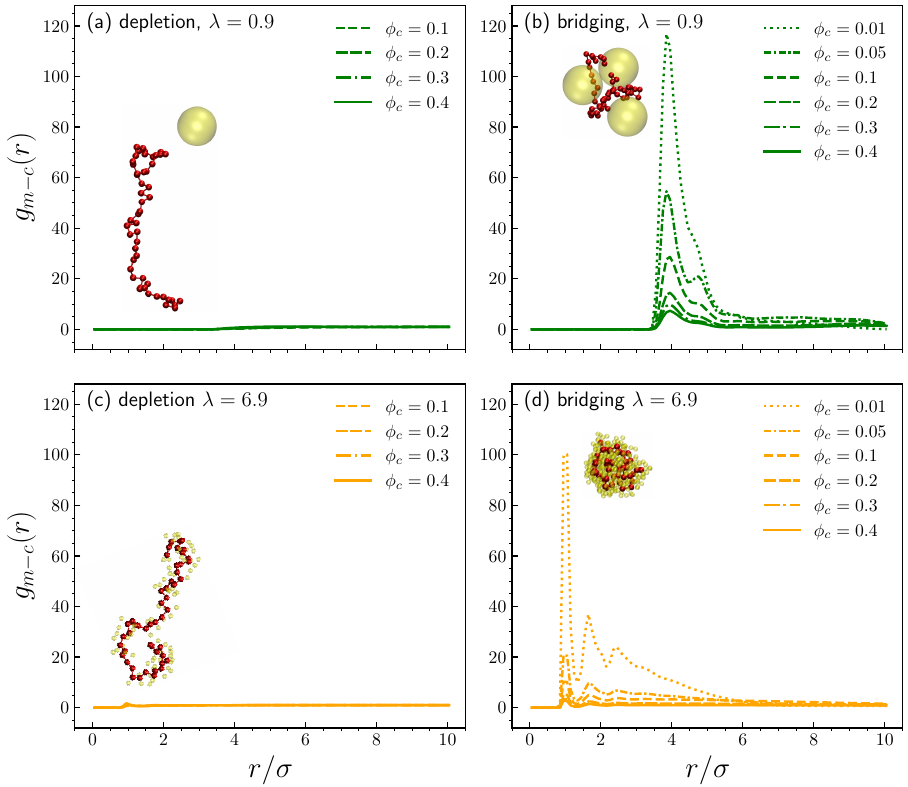}
\caption{Monomer-crowder pair distribution functions $g_{m\text{-}c}(r)$ for depletion and bridging regimes at two crowder size ratios. (a)~Depletion, $\lambda = 0.9$; (b)~Bridging, $\lambda = 0.9$; (c)~Depletion, $\lambda = 6.9$; (d)~Bridging, $\lambda = 6.9$ for a range of crowder density values.  Insets show representative simulation snapshots at the lowest $\phi_c$ studied, illustrating the contrasting adsorption geometries for large ($\lambda = 0.9$) and small ($\lambda = 6.9$) crowders.}
\label{fig:mc_gr}
\end{figure}

To probe the microscopic basis of polymer collapse and reentrant expansion, we computed the monomer-crowder pair distribution function $g_{m\text{-}c}(r)$ for representative crowder sizes. Figure~\ref{fig:mc_gr} compares depletion and bridging regimes for size ratios $\lambda=0.9$ and $\lambda=6.9$ for different crowder densities.

In the depletion case [Figs.~\ref{fig:mc_gr}(a,c)], $g_{m\text{-}c}(r)$ remains essentially flat and close to unity at all crowder densities $\phi_c$ that reflects the steric exclusion of crowders from the monomer surface. The absence of any pronounced adsorption peak confirms that crowders do not preferentially bind to the chain; they are simply excluded from the polymer domain by hard-core repulsion, consistent with the Asakura-Oosawa depletion picture~\cite{Kang2015PRL}. This microscopic picture is consistent with the monotonic compaction observed in the depletion regime (Fig.~\ref{rg-wca-bridge}): crowders act on the chain from the outside via osmotic pressure, never accumulating within it.

When monomer-crowder attractions are introduced [Figs.~\ref{fig:mc_gr}(b,d)], $g_{m\text{-}c}(r)$ develops a pronounced contact peak at $r \simeq \sigma_{mc}$, signaling strong adsorption of crowders onto the polymer chain. At the lowest $\phi_c$ studied, the contact peak heights are comparable for both crowder sizes, but the adsorption geometry differs. For large crowders ($\lambda = 0.9$) [Fig.~\ref{fig:mc_gr}(b)], the peak is broader: as the simulation snapshots illustrate, the polymer wraps around the crowder surface. Although the number of crowders is smaller for $\lambda=0.9$, the comparable height of the contact peak suggests that each large crowder can simultaneously interact with many monomers by allowing the polymer to wrap around its surface. For small crowders ($\lambda = 6.9$) [Fig.~\ref{fig:mc_gr}(d)], the peak is sharper, reflecting the more localized geometry in which small crowders sit between monomers along the chain, simultaneously contacting multiple chain segments. It is this multivalent contact geometry that drives the cooperative collapse seen in Fig.~\ref{rg-wca-bridge}(b) at low $\phi_c$.

As $\phi_c$ increases in the bridging regime, the contact peak height decreases systematically for both crowder sizes, and for $\lambda = 6.9$ the peak nearly vanishes by $\phi_c = 0.3$-$0.4$, with $g_{m\text{-}c}(r)$ approaching the flat profile of the depletion case. Part of this decrease reflects the normalization inherent in $g(r)$: because $g_{m\text{-}c}(r) = \rho_{\mathrm{local}}(r)/\rho_{\mathrm{bulk}}$, the peak height drops as the bulk crowder density $\rho_{\mathrm{bulk}}$ grows with $\phi_c$, even if the local adsorbed population remains constant. The absolute number of neighboring crowders $n_{\mathrm{c}}$ (Sec.~\ref{subsection:neighbor}) confirms that crowders are not expelled from the chain at high $\phi_c$; rather, what changes is the adsorption geometry. At low $\phi_c$, each adsorbed crowder finds multiple unoccupied monomers nearby and bridges between them. At high $\phi_c$, most monomer sites are already occupied by previously adsorbed crowders, so incoming crowders can only form short-range, single-monomer contacts that contribute no bridging connectivity. In this sense, monomers become locally passivated: the bridging valency per crowder decreases from multivalent to effectively monovalent, suppressing the intrachain crosslinks that stabilize the collapsed globule. A secondary peak visible at $r \approx 2\sigma_{mc}$ at intermediate densities in panel (d) indicates the onset of a second coordination shell of crowders around the chain, consistent with dense, uniform coating of the polymer surface. In both cases, the transition from a high contact-peak regime to a low contact-peak regime in $g_{m\text{-}c}(r)$ maps onto the coil-to-globule and globule-to-coil transitions seen in Fig.~\ref{rg-wca-bridge}(b), providing microscopic corroboration of the bridging-saturation mechanism.

\subsection{Effect of monomer-crowder interactions on evolution of  neighbor crowders}\label{subsection:neighbor}

\begin{figure}
\centering
\includegraphics[width=0.95\columnwidth]{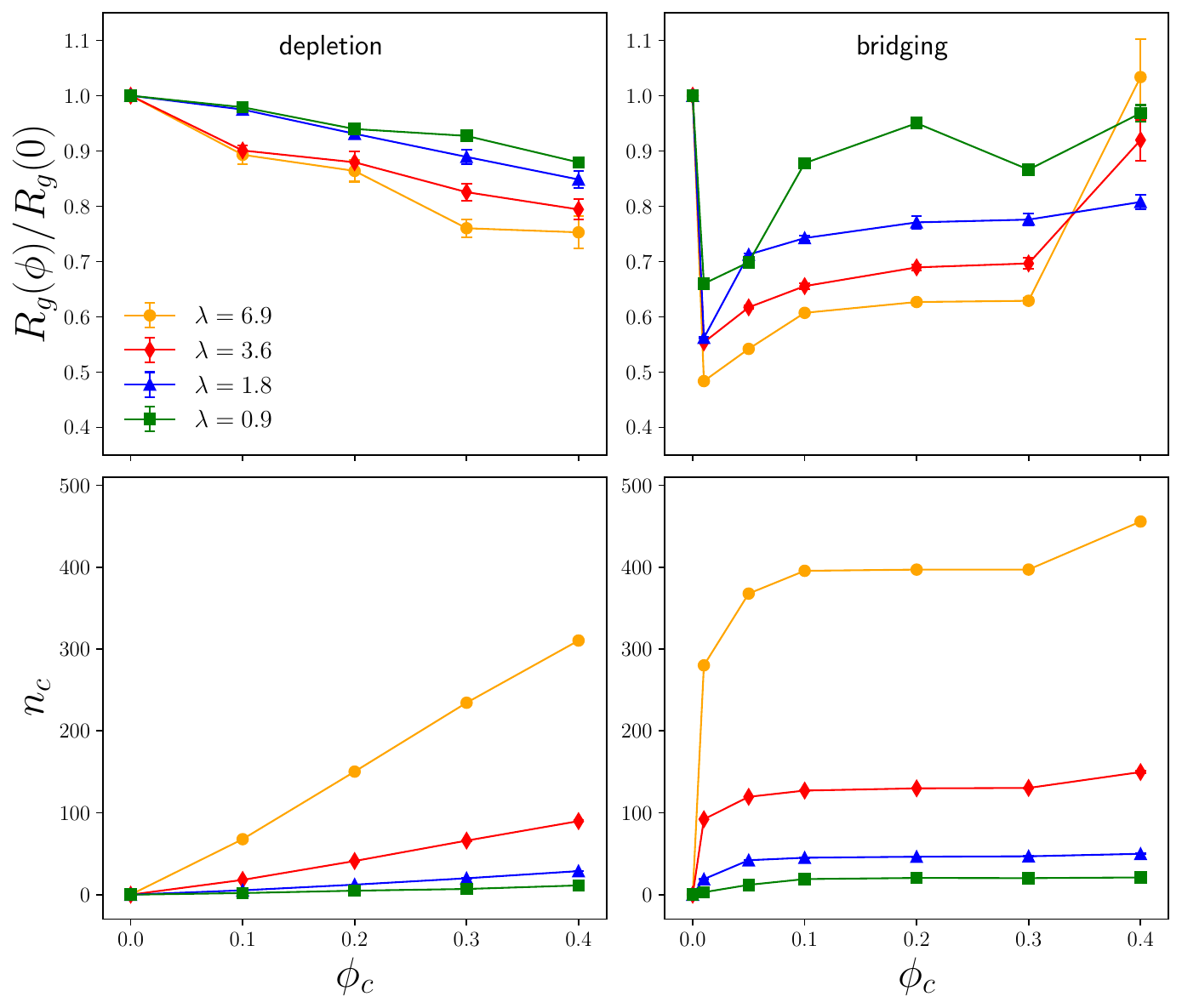}
\caption{Number of neighboring crowders $n_\mathrm{c}$ within $1.5\,\sigma_{mc}$ of any monomer (bottom panels) and normalized radius of gyration $R_g(\phi_c)/R_g(0)$ (top panels) as a function of crowder volume fraction $\phi_c$, for depletion (left) and bridging (right) regimes at four size ratios $\lambda$.}
\label{fig:noc}
\end{figure}

\begin{figure}
\centering
\includegraphics[width=0.95\columnwidth]{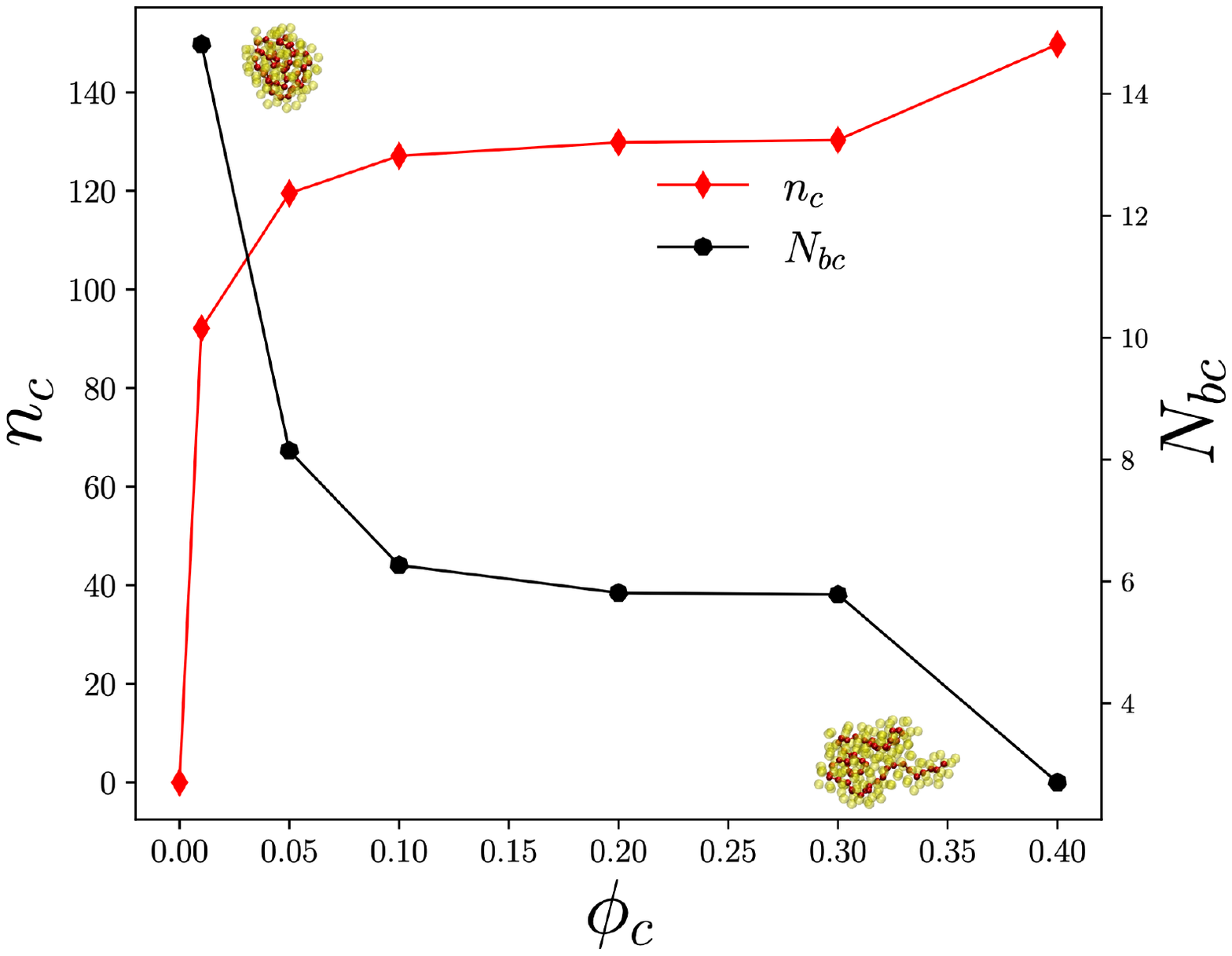}
\caption{Total number of neighboring crowders $n_c$ (red, left axis) and number of bridging crowders $N_{bc}$ (black, right axis; see Eq.~\ref{eq:bridging_crowders}) as a function of $\phi_c$ for $\lambda = 6.9$. While $n_c$ plateaus at high $\phi_c$, $N_{bc}$ decreases monotonically and approaches zero, indicating that crowders remain adsorbed on the chain but lose their bridging connectivity.}
\label{fig:nocBridge}
\end{figure}

To further probe the bridging-saturation mechanism, we computed the number of neighboring crowders, $n_\mathrm{c}$, within a distance of $1.5\,\sigma_{mc}$ from any monomer bead, and tracked its evolution with $\phi_c$ alongside $R_g(\phi_c)/R_g(0)$. Figure~\ref{fig:noc} shows both quantities for depletion (left column) and bridging (right column).

\textit{Depletion regime.} In the purely repulsive case [Fig.~\ref{fig:noc}, bottom-left], $n_\mathrm{c}$ increases linearly with $\phi_c$ for all size ratios, with smaller crowders ($\lambda = 6.9$) contributing more neighbors at fixed $\phi_c$ simply because they are more numerous at the same volume fraction. This linear growth reflects the fact that, in the absence of adsorption, the local crowder density near the chain tracks the bulk density with no preferential accumulation at the monomer surface. $n_\mathrm{c}$ grows linearly while $R_g$ decreases monotonically [Fig.~\ref{fig:noc}, top-left], confirming that chain compaction in the depletion regime is driven by bulk osmotic pressure rather than by local crowder accumulation.

\textit{Bridging regime.} The bridging case [Fig.~\ref{fig:noc}, bottom-right] is qualitatively different. For all $\lambda$, $n_\mathrm{c}$ rises steeply and non-linearly at low crowder densities, then saturates to an approximately constant plateau over a broad range of $\phi_c$. For $\lambda = 6.9$, $n_\mathrm{c}$ reaches $\sim\!280$ at $\phi_c = 0.01$ and saturates near $400$ by $\phi_c \approx 0.05$-$0.1$, remaining essentially flat through $\phi_c = 0.3$. At $\phi_c = 0.1$, the bridging case shows $n_\mathrm{c} \approx 370$ for $\lambda = 6.9$, compared to $\sim\!70$ in the depletion case, a factor of $\sim\!5$ excess that reflects preferential adsorption driven by monomer-crowder attraction. The steep initial rise in $n_\mathrm{c}$ at low $\phi_c$ corresponds to the regime of rapid collapse in $R_g$ [Fig.~\ref{fig:noc}, top-right], while the subsequent plateau in $n_\mathrm{c}$ corresponds to the regime of gradual reentrant expansion: monomer binding sites become occupied, and additional crowders can no longer form new bridging contacts.

The $n_\mathrm{c}$ plateau confirms that crowders remain adsorbed on the chain at high $\phi_c$, but it does not by itself reveal whether those adsorbed crowders retain their bridging connectivity. To address this, we define a \emph{bridging crowder} as one that simultaneously interacts with at least $k = 6$ monomers within a sphere of radius $1.5\,\sigma_{mc}$, following the criterion introduced in Ref.~\cite{Garg2023JCP}. The total number of bridging crowders is
\begin{equation}
N_{bc} = \sum_{k=6}^{\infty} n_k\,,
\label{eq:bridging_crowders}
\end{equation}
where $n_k$ is the number of crowders with exactly $k$ monomers within the cutoff distance.

Figure~\ref{fig:nocBridge} shows $n_c$ and $N_{bc}$ as a function of $\phi_c$ for $\lambda = 6.9$. While the total number of neighboring crowders $n_c$ plateaus and remains high across the full density range, $N_{bc}$ decreases monotonically from $\sim\!15$ at $\phi_c = 0.01$ to near zero by $\phi_c = 0.4$. This contrast strongly suggests that the crowders do not leave the chain as $\phi_c$ increases, but they cease to bridge. At low $\phi_c$, each adsorbed crowder finds multiple unoccupied monomers nearby and bridges between them. At high $\phi_c$, most monomer sites are already occupied by other crowders, so each crowder can contact only one or a few monomers, falling below the bridging threshold. The intrachain crosslinks that stabilize the collapsed globule are progressively lost, and the chain reexpands.

\subsection{Chain statistics: universality in depletion versus breakdown in the bridging regime}\label{subsection:chainstatneutral}
\begin{figure}
\centering
\includegraphics[width=0.95\columnwidth]{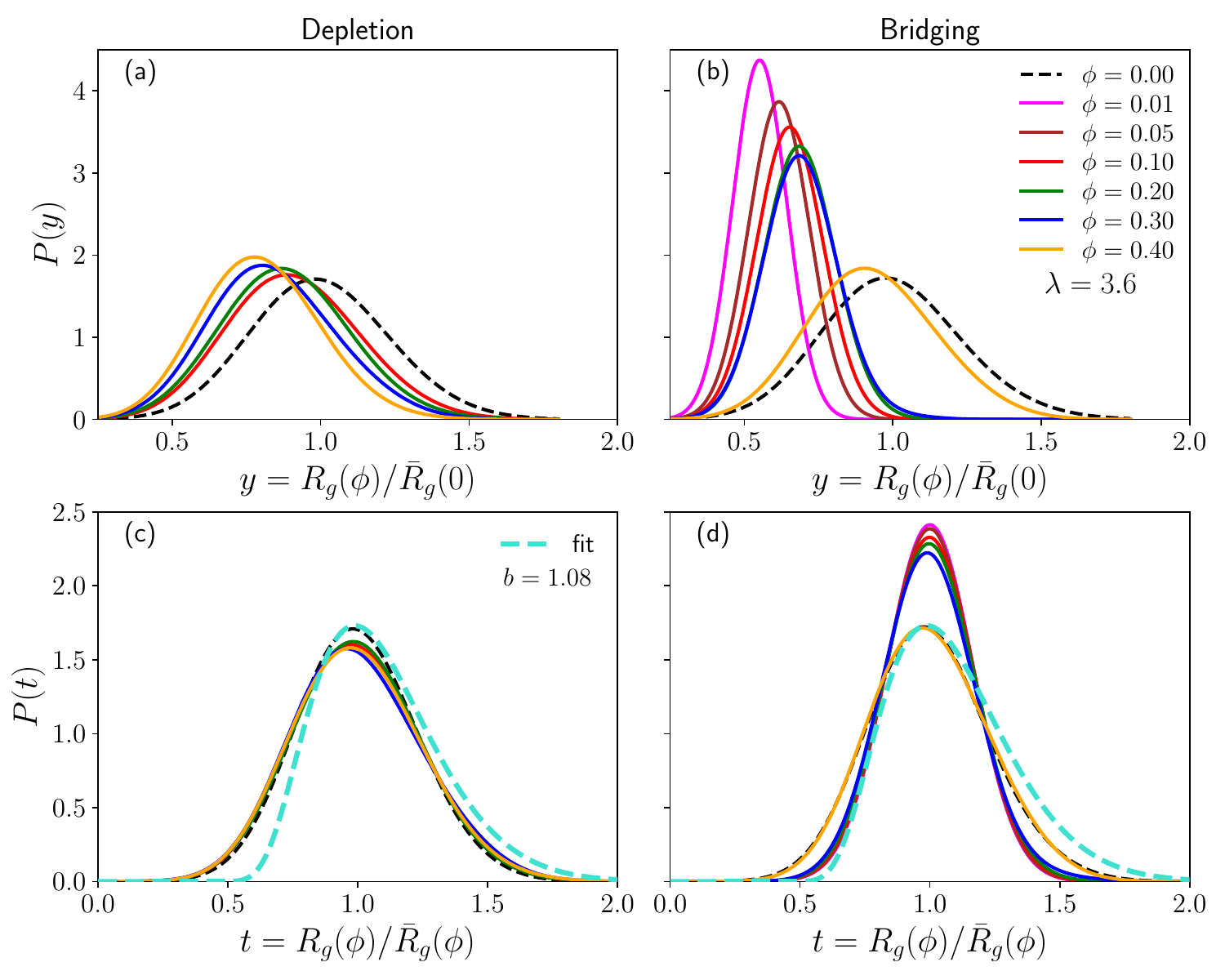}
\caption{Distributions of the normalized radius of gyration for $\lambda = 3.6$ in the depletion (left) and bridging (right) regimes at $\phi_c = 0$--$0.4$. (a,b)~$P(y)$ with $y = R_g(\phi_c)/\bar{R}_g(0)$. (c,d)~$P(t)$ with $t = R_g(\phi_c)/\bar{R}_g(\phi)$, used to assess universality of chain statistics. Dashed line in (c): SAW master curve [Eq.~(\ref{eq:Pt}), $b = 1.08$].}
\label{fig:Pgr}
\end{figure}

A central result of Kang \emph{et al.}~\cite{Kang2015PRL} for purely repulsive crowders was that the distribution of $R_g$ values, when rescaled by the mean $\bar{R}_g(\phi)$, collapses onto a single universal curve independent of both $\phi$ and $\lambda$: the distribution of a self-avoiding walk (SAW). This universality implies that depletion-driven compaction does not alter the statistical character of the chain; it merely rescales the typical size while preserving SAW-like conformational fluctuations. We now examine whether this universality survives in the bridging regime.

Figure~\ref{fig:Pgr} shows the distribution $P(y)$ of the normalized radius of gyration $y = R_g(\phi_c)/\bar{R}_g(0)$ [panels (a,b)], and the rescaled distribution $P(t)$ with $t = R_g(\phi_c)/\bar{R}_g(\phi)$ [panels (c,d)], for $\lambda = 3.6$ in both depletion and bridging regimes at a range of $\phi_c$.

\textit{Depletion regime.} In the depletion case [Fig.~\ref{fig:Pgr}(a)], the distributions $P(y)$ are unimodal at all $\phi_c$, with a single peak that shifts progressively leftward from $y \approx 1.0$ at $\phi_c = 0$ to $y \approx 0.85$--$0.88$ at $\phi_c = 0.4$, reflecting gradual, continuous compression with increasing crowder density. When these distributions are rescaled by $\bar{R}_g(\phi)$ to give $P(t)$ [Fig.~\ref{fig:Pgr}(c)], all curves at different $\phi_c$ collapse onto a single universal master curve, well described by the Lhuillier-de~Gennes SAW distribution~\cite{Kang2015PRL,DeGennes1979,Lhuillier1988JPhys}:
\begin{equation}
P(t) = \mathcal{N}\, e^{-(bt)^{-15/4} - (bt)^{5/2}},
\label{eq:Pt}
\end{equation}
with $b = 1.08$ (compared to $b = 1.12$ reported in Kang \emph{et al.} for $N = 100$; the small difference is attributable to the shorter chain length $N = 50$ used here). This collapse demonstrates that the chain retains SAW statistics throughout the depletion regime: crowding rescales the mean size but does not alter the universality class of the conformational fluctuations.

\textit{Bridging regime.} The bridging case is qualitatively different. In $P(y)$ [Fig.~\ref{fig:Pgr}(b)], at $\phi_c = 0$ the distribution is a single peak at $y \approx 1.0$, identical to the free chain. At $\phi_c = 0.01$, the peak shifts to $y \approx 0.5$--$0.6$ and broadens considerably, indicating that the chain has collapsed into a globule-like state. The distribution remains unimodal at all $\phi_c$, but the shift between $\phi_c = 0$ and $\phi_c = 0.01$ is large and abrupt compared to the gradual leftward drift seen in the depletion case, reflecting the cooperative nature of bridging-induced collapse. By $\phi_c = 0.05$--$0.10$, the peak narrows near $y \approx 0.6$, confirming that the chain is predominantly in the collapsed bridged-globule state. As $\phi_c$ increases further to $0.20$--$0.40$, the peak shifts back toward $y \approx 0.9$--$1.0$, tracing the complete coil-globule-coil reentrant trajectory in the conformational distributions.

The rescaled distributions $P(t)$ in the bridging regime [Fig.~\ref{fig:Pgr}(d)] do not collapse onto a universal curve. At low $\phi_c$ ($0.01$--$0.05$), $P(t)$ is broad and shifted relative to the SAW master curve, reflecting the non-SAW statistics of the bridged globule. At high $\phi_c$ ($0.30$--$0.40$), $P(t)$ begins to approach the SAW form again, consistent with the reentrant expansion restoring coil-like statistics. The failure of the rescaled distributions to collapse onto a single master curve shows that bridging does not merely rescale the chain size; it changes the universality class of the conformational ensemble. The chain passes through at least three statistically distinct regimes as $\phi_c$ increases: SAW coil at $\phi_c = 0$, a non-universal bridged globule at intermediate $\phi_c$, and a partially recovered SAW-like state at high $\phi_c$.

Table~\ref{tbl:dimless_parameter_x} provides a quantitative comparison between the bridging and depletion regimes. In the depletion framework of Kang \emph{et al.}~\cite{Kang2015PRL}, cooperative collapse requires $x \gg 1$. In our depletion simulations, this condition is approached only for $\lambda = 6.9$ at $\phi_c \geq 0.1$ ($x \approx 2.0$--$3.2$), never exceeds $x \approx 1.65$ for $\lambda = 3.6$, and remains below $0.82$ for $\lambda \leq 1.8$ across all densities. Consistent with this, depletion produces only modest, monotonic compaction for all $\lambda$ [Fig.~\ref{rg-wca-bridge}(a)]. In the bridging regime, by contrast, the sharpest collapse occurs at $\phi_c = 0.01$, where $x$ ranges from $0.12$ ($\lambda = 0.9$) to $0.92$ ($\lambda = 6.9$), values at which depletion-driven collapse is negligible. Bridging thus operates in a regime inaccessible to depletion: it does not require $x \gg 1$, but instead exploits the adsorption affinity of even a small number of crowders to generate multivalent intrachain crosslinks. The two mechanisms are distinct in character: depletion is osmotic and requires $x \gg 1$; bridging is connective and operates at any $x$.

Taken together, the results of this section establish a clear contrast between the two crowder regimes. Depletion preserves SAW universality at all densities, consistent with the purely osmotic picture of Kang \emph{et al.}~\cite{Kang2015PRL}. Bridging breaks this universality through a sharp, cooperative coil-globule transition at low $\phi_c$, drives the chain through non-SAW globule statistics at intermediate $\phi_c$, and only partially restores SAW behavior at high $\phi_c$ as bridging valency is saturated. The partial recovery of SAW-like statistics at high $\phi_c$ confirms that the reexpanded state produced by bridging saturation is conformationally similar to the original free chain.

\subsection{Charged polymer with monovalent counterions: persistence and amplification of reentrant behavior}

\begin{figure}
\centering
\includegraphics[width=0.9\columnwidth]{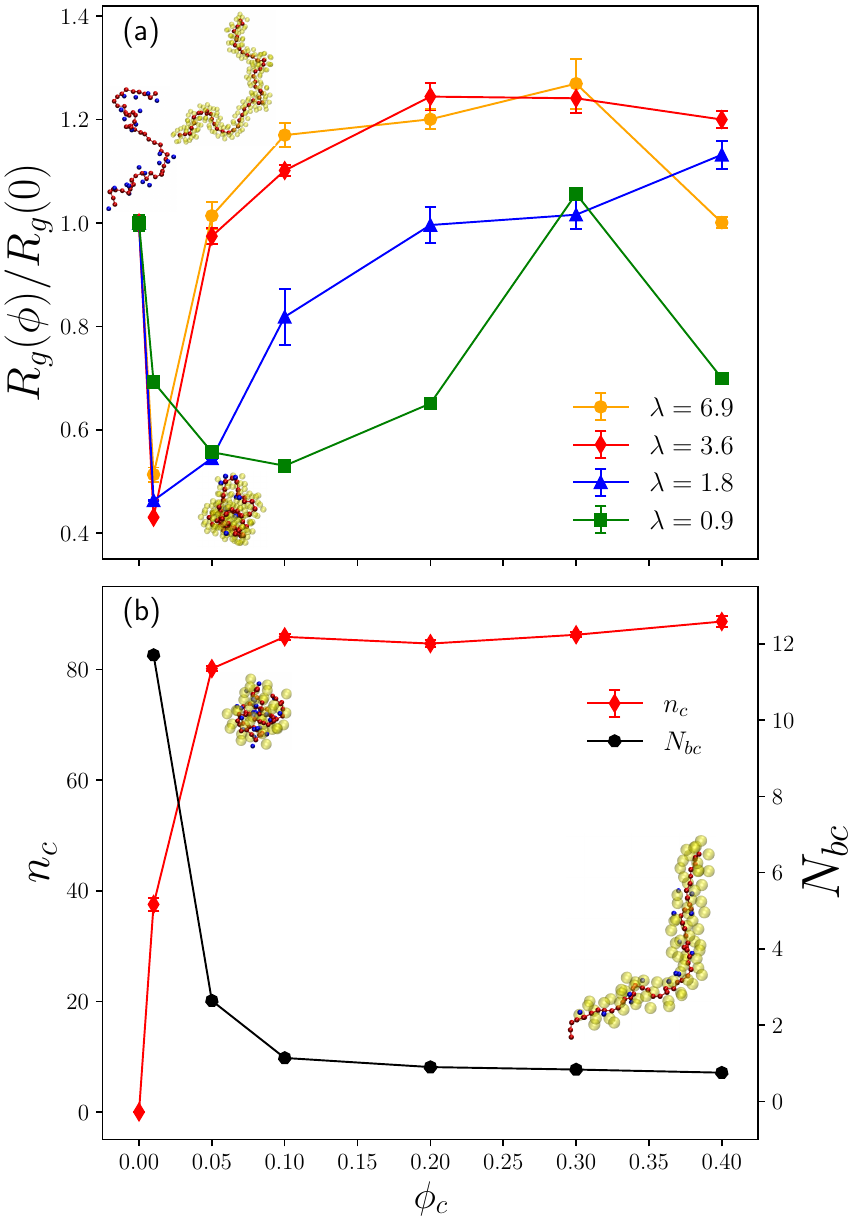}
\caption{(a)~Normalized radius of gyration $R_g(\phi_c)/R_g(0)$ of a uniformly charged polymer neutralized by monovalent counterions ($Z = 1$) in the bridging regime, for four size ratios $\lambda = R_g(0)/\sigma_c$. (b)~Total number of neighboring crowders $n_c$ (red, left axis) and number of bridging crowders $N_{bc}$ (black, right axis) as a function of $\phi_c$ for $\lambda = 3.6$. Insets show representative simulation snapshots at low and high $\phi_c$. Error bars: standard errors from block averaging.}
\label{fig:charged_rg}
\end{figure}

Having established the bridging-saturation mechanism for neutral polymers, we now ask whether it survives when the polymer carries backbone charge. Our earlier work has shown that attractive crowders can collapse charged chains even in the presence of intrachain Coulomb repulsion~\cite{Tripathi2023JCP}. What has not been examined is whether the density-driven reentrant expansion observed for neutral chains persists in charged systems, and if so, how electrostatics modifies the amplitude and $\lambda$-dependence of the reentrant response. We consider a uniformly charged polymer neutralized by explicit monovalent counterions ($Z = 1$); results for trivalent counterions ($Z = 3$) are presented in the Supplemental Material and summarized at the end of this section. The crowders remain neutral and interact either purely repulsively (depletion regime) or attractively with monomers (bridging regime). The free-chain size for $Z = 1$ is $R_g(0) = 8.22\sigma$, compared to $5.17\sigma$ for neutral chains (Table~S1).

\textit{Depletion regime.} When crowders interact purely via excluded volume, the charged polymer exhibits weak, monotonic compaction with increasing $\phi_c$, qualitatively identical to the neutral depletion case. The presence of counterions and backbone charge does not alter the depletion response; steric osmotic pressure alone produces monotonic compression. Figure S1 provides a side-by-side comparison of $R_g(\phi_c)/R_g(0)$ across neutral, $Z=1$, and $Z=3$ chains in both regimes, confirming that all three systems show monotonic compaction under depletion, with the charged chains somewhat more resistant to compaction due to intrachain Coulomb repulsion.

\textit{Bridging regime.} Figure~\ref{fig:charged_rg}(a) shows $R_g(\phi_c)/R_g(0)$ for the bridging regime with monovalent counterions. The polymer collapses at low $\phi_c \approx 0.01$, with $R_g(\phi_c)/R_g(0)$ dropping to $\sim 0.46$--$0.73$ depending on $\lambda$, comparable in depth and onset to the neutral bridging case. This indicates that the monomer-crowder attraction ($\epsilon_{mc} = 4\,k_BT$) is strong enough to overcome intrachain Coulomb repulsion even at low crowder density.

The reexpansion at high $\phi_c$ is present for all $\lambda$, but its magnitude is considerably larger than in the neutral case. For $\lambda = 3.6$, the chain rises from a minimum of $\sim 0.69$ at $\phi_c = 0.01$ to $R_g(\phi_c)/R_g(0) \approx 2.0$ at $\phi_c = 0.20$--$0.30$, expanding to twice its free-solution size. Similarly, $\lambda = 1.8$ reaches $R_g(\phi_c)/R_g(0) \approx 1.82$ at $\phi_c = 0.40$. The chain thus expands well beyond its unperturbed size upon bridging saturation, in contrast to the neutral case where $R_g(\phi_c = 0.40)$ approximately recovers to $R_g(0)$.

The $\lambda$-dependence of the reexpansion is different from the neutral case, with regards to the crowder size. For neutral chains, the smallest crowders (largest $\lambda$) produced the most complete recovery. Here, the largest overshoot is produced by intermediate crowder sizes ($\lambda = 3.6$ and $\lambda = 1.8$), while the smallest crowders ($\lambda = 6.9$) show a more modest reexpansion ($\sim 1.25$) and the largest crowders ($\lambda = 0.9$) show only weak reexpansion. This difference arises from the interplay between bridging and electrostatics. For small crowders ($\lambda = 6.9$), bridging is efficient and the chain collapses deeply, but at high $\phi_c$ the dense coating of adsorbed small crowders partially screens the backbone charge, limiting the electrostatic contribution to reexpansion. For intermediate crowder sizes, bridging is still effective at low $\phi_c$ but saturates less completely at high $\phi_c$, leaving a larger fraction of the backbone charge exposed to drive electrostatic swelling once bridging connectivity is lost. For large crowders ($\lambda = 0.9$), both bridging and saturation are weak, and the reexpansion is correspondingly modest. 

The likely physical picture is as follows. During the bridged-globule phase, bridging contacts crosslink the chain and suppress long-range Coulomb repulsion by bringing distant charged monomers into close proximity, analogous to charge neutralization by multivalent counterions in polyelectrolyte condensation~\cite{Raspaud1998BJ,Olvera1995JCP}. As $\phi_c$ increases and bridging valency is saturated, these crosslinks dissolve and the backbone charges, which have been constrained in proximity throughout the collapsed phase, experience renewed mutual repulsion that drives the chain well beyond its original size. In this sense, the collapsed bridging phase acts as a conformational capacitor: electrostatic repulsion energy is stored during collapse and released upon saturation.

\textit{Bridging crowder analysis.} Figure~\ref{fig:charged_rg}(b) shows the total number of neighboring crowders $n_c$ and the number of bridging crowders $N_{bc}$ (Eq.~\ref{eq:bridging_crowders}) as a function of $\phi_c$ for $\lambda = 3.6$. The qualitative pattern is the same as for neutral chains (Fig.~\ref{fig:nocBridge}): $n_c$ rises and plateaus while $N_{bc}$ decreases monotonically, confirming that the bridging-saturation mechanism operates in the charged system as well. There are, however, two quantitative differences. First, the $n_c$ plateau for the charged chain ($\sim 85$) is substantially lower than for neutral chains ($\sim 400$ at the same $\lambda$), reflecting the more extended monomer distribution in the charged chain, which reduces local monomer density and hence crowder packing. Second, despite the lower absolute number of adsorbed crowders, $N_{bc}$ starts at a comparable value ($\sim 12$ vs $\sim 15$ for neutral) and drops to near zero over a similar range of $\phi_c$, indicating that the bridging connectivity per adsorbed crowder is similar in both cases. The loss of bridging connectivity thus proceeds on the same density scale regardless of backbone charge; what differs is the consequence of that loss, since the charged chain has stored electrostatic energy during the collapsed phase that is released as super-swelling upon saturation.

\textit{Counterion displacement.} The counterion neighbor count $n_\mathrm{ci}$ (Fig. S2, bottom panels) provides additional support for the conformational capacitor picture. In depletion, $n_\mathrm{ci}$ remains near its baseline value of $\sim 26$ at all $\phi_c$ and $\lambda$, confirming that repulsive crowders do not perturb the counterion cloud. In bridging, the behavior depends on $\lambda$. For small crowders ($\lambda = 6.9$), $n_\mathrm{ci}$ drops from $\sim 26$ at $\phi_c = 0$ to nearly zero by $\phi_c = 0.05$ and remains near zero throughout the reentrant expansion. This shows that small bridging crowders displace counterions from the monomer surface by competing for adsorption sites. The counterions do not reassociate with the chain even as it expands at high $\phi_c$, so the backbone charges remain unscreened, providing the electrostatic driving force for the large overshoot. This also explains why $\lambda = 6.9$ produces the deepest collapse but only moderate normalized reexpansion: the dense small-crowder coating expels counterions but also partially suppresses electrostatic expansion sterically. For large crowders ($\lambda = 0.9$), $n_\mathrm{ci}$ remains near its baseline value throughout, indicating that large crowders cannot efficiently displace counterions. The counterion cloud remains intact, partially screening the backbone charge and limiting the reexpansion. The monomer-crowder pair distribution functions $g_{m\text{-}c}(r)$ for $Z = 1$ (Fig. S3) closely parallel the neutral chain results (Fig.~\ref{fig:mc_gr}), confirming that the microscopic adsorption picture is qualitatively preserved in the presence of backbone charge.

\textit{Trivalent counterions.} Qualitatively similar behavior is observed with $Z = 3$ counterions (Figs. S4 and S5). The $n_\mathrm{c}$ plateau in bridging persists, and a reentrant transition is observed, confirming that the saturation mechanism operates regardless of counterion valency. However, counterion displacement is less complete with trivalent counterions: $n_\mathrm{ci}$ drops but does not reach zero, consistent with stronger counterion condensation at higher valency. The partial retention of the counterion cloud screens the backbone charge more effectively during reexpansion, contributing to the more moderate normalized overshoot ($R_g(\phi_c)/R_g(0) \approx 1.22$ for $\lambda = 3.6$). The free-chain size for $Z = 3$ ($R_g(0) = 6.55\sigma$) is also smaller than for $Z = 1$ ($8.22\sigma$) due to stronger counterion condensation, which further reduces the normalized ratio for comparable absolute expansion.

These results show that reentrant expansion is not suppressed by backbone electrostatics. Coulomb repulsion acts as an amplifier of the non-monotonic response: it is suppressed during the bridged-globule phase and reasserts itself once bridging connectivity is lost at high $\phi_c$. The qualitative sequence coil $\rightarrow$ bridged globule $\rightarrow$ super-swollen coil is present for both counterion valencies, with electrostatics reshaping the amplitude but not the topology of the reentrant transition.

\subsection{Chain statistics of charged polymers: super-SAW regime}

\begin{figure}
\centering
\includegraphics[width=0.9\columnwidth]{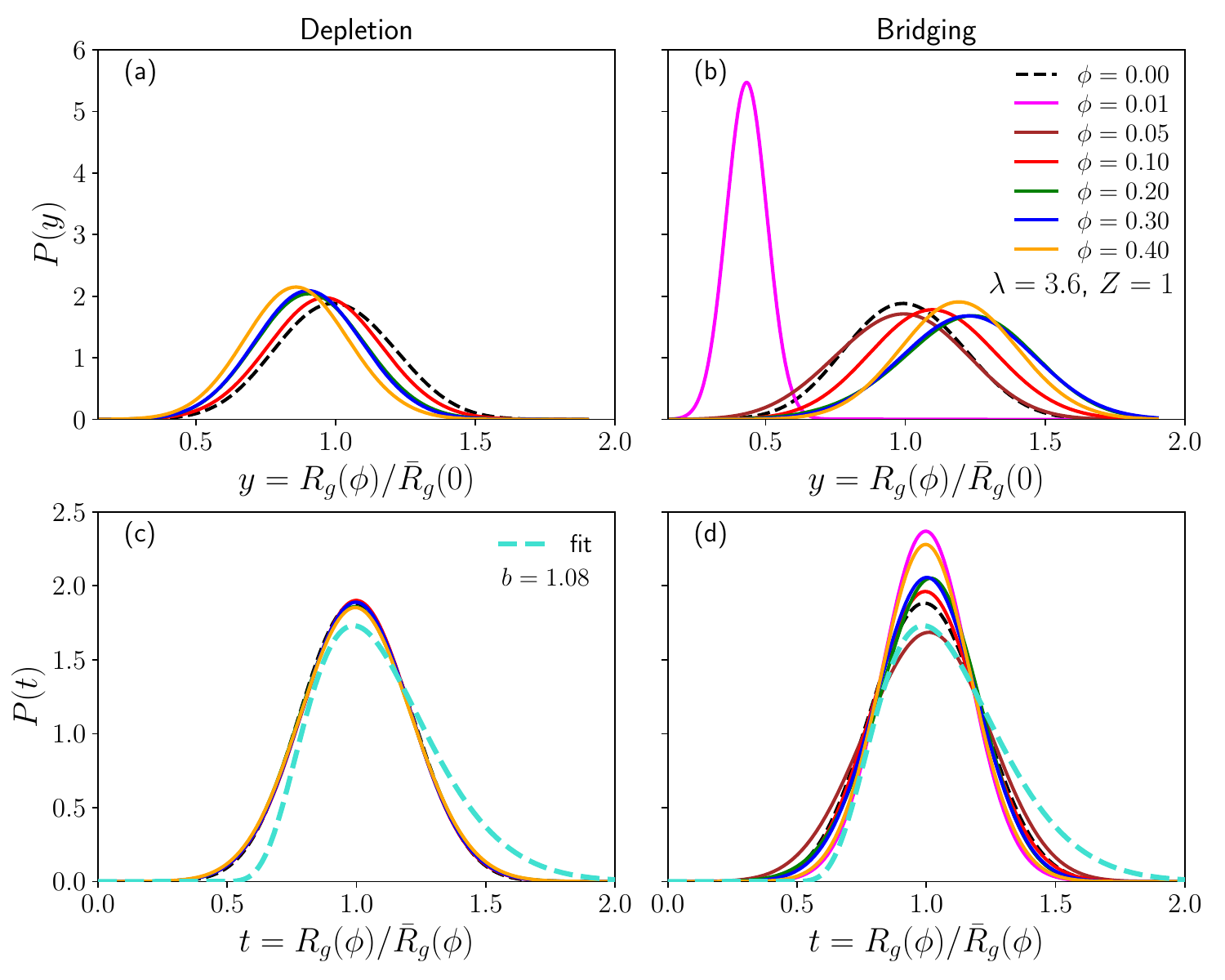}
\caption{Distributions of the normalized radius of gyration for $\lambda = 3.6$, $Z = 1$ in the depletion (left) and bridging (right) regimes at $\phi_c = 0$--$0.4$, parallel to Fig.~\ref{fig:Pgr} for neutral chains. (a,b)~$P(y)$ with $y = R_g(\phi)/\bar{R}_g(0)$; (c,d)~$P(t)$ with $t = R_g(\phi)/\bar{R}_g(\phi)$. Dashed line in (c): SAW master curve ($b = 1.08$).}
\label{fig:charged_Pgr}
\end{figure}

Having established that bridging breaks SAW universality in neutral chains, we now examine whether the same characterization applies to charged polymers, and whether the super-swelling seen in $R_g(\phi_c)/R_g(0)$ for $Z = 1$ is reflected in the chain statistics. Figure~\ref{fig:charged_Pgr} shows $P(y)$ and $P(t)$ for $\lambda = 3.6$, $Z = 1$, parallel to Fig.~\ref{fig:Pgr} for the neutral case.

\textit{Depletion regime.} In the depletion case [Fig.~\ref{fig:charged_Pgr}(a)], $P(y)$ shifts progressively leftward with increasing $\phi_c$, confirming monotonic compaction. The distributions are narrower and taller than in the neutral case, reflecting the stiffening effect of intrachain Coulomb repulsion, which restricts the chain to a narrower range of conformations near its electrostatically extended equilibrium. The leftward shift with increasing $\phi_c$ is slightly smaller than for neutral chains, consistent with electrostatic repulsion opposing depletion-driven compaction. Despite these quantitative differences, the rescaled distributions $P(t)$ [Fig.~\ref{fig:charged_Pgr}(c)] collapse onto the same SAW master curve as the neutral case, with the same fitted parameter $b = 1.08$. SAW universality in the depletion regime is thus robust to backbone charge: Coulomb repulsion modifies the mean size and the width of the distribution but does not alter the universality class of the conformational fluctuations.

\textit{Bridging regime.} The bridging case [Fig.~\ref{fig:charged_Pgr}(b,d)] differs from the neutral counterpart in two respects: the depth of the initial collapse and the extent of the subsequent reexpansion. In $P(y)$ [Fig.~\ref{fig:charged_Pgr}(b)], the distributions are unimodal at all $\phi_c$, as in the neutral case, but the reentrant trajectory spans a wider range. The peak shifts from $y \approx 1.0$ at $\phi_c = 0$ to $y \approx 0.5$ at $\phi_c = 0.01$ (collapse), then back to $y \approx 1.2$--$1.3$ at $\phi_c = 0.20$--$0.40$, well beyond the free-chain value. This super-swelling, absent in neutral chains, reflects the restored electrostatic repulsion between backbone charges upon bridging saturation. The distributions at high $\phi_c$ are also broader than at $\phi_c = 0$, indicating larger conformational fluctuations in the super-swollen state. The rescaled distributions $P(t)$ [Fig.~\ref{fig:charged_Pgr}(d)] confirm this picture. At low $\phi_c$, $P(t)$ shifts to the left of the SAW master curve, as in the neutral case. At high $\phi_c$, however, $P(t)$ shifts to the \emph{right} of the master curve, with weight at $t > 1$ exceeding that expected for a self-avoiding chain. We term this regime \emph{super-SAW}: the chain is more extended than a SAW of equivalent contour length, driven by long-range electrostatic repulsion restored once bridging connectivity is lost. No analog of this regime exists for neutral chains, where $P(t)$ at high $\phi_c$ recovers toward but does not exceed the SAW master curve.

The conformational statistics of the $Z = 1$ chain thus trace a four-regime sequence as $\phi_c$ increases: (i)~a SAW coil at $\phi_c = 0$, with $P(t)$ on the master curve; (ii)~a bridged globule at $\phi_c \approx 0.01$--$0.10$, with $P(y)$ shifted to low $y$ and $P(t)$ shifted left of the master curve; (iii)~an intermediate reentrant coil at $\phi_c \approx 0.10$--$0.20$, with partial statistical recovery; and (iv)~a super-SAW regime at $\phi_c \gtrsim 0.20$, with $P(y)$ centered at $y > 1$ and $P(t)$ shifted right of the master curve. This four-regime sequence contrasts with the three-regime sequence (SAW $\to$ bridged globule $\to$ partially recovered SAW) observed for neutral chains. The super-SAW regime is the statistical signature of electrostatic amplification of the reentrant transition.

The corresponding distributions for $Z = 3$ (Fig. S6) show qualitatively similar features: SAW universality is preserved under depletion, universality breaks down under bridging, and a super-SAW regime is present at high $\phi_c$. The rightward shift of $P(t)$ at high $\phi_c$ is less pronounced for $Z = 3$ than for $Z = 1$, consistent with the smaller normalized overshoot in $R_g(\phi_c)/R_g(0)$ for the more strongly charged chain (Fig.~\ref{fig:charged_rg}b) and the incomplete counterion displacement discussed in the previous section.

\section{Discussion and Conclusion}\label{Sec-4}

The central finding of this work is that a single class of attractive crowders, interacting with a homopolymer at fixed interaction strength, produces a complete density-driven reentrant coil-globule-coil transition as crowder volume fraction $\phi_c$ is varied. The mechanism is saturable geometric bridging: at low $\phi_c$, each crowder simultaneously contacts multiple monomers and generates intrachain crosslinks that drive collapse; at high $\phi_c$, monomer binding sites become occupied, bridging valency drops (as confirmed by the decrease in $N_{bc}$, Fig.~\ref{fig:nocBridge}), and the chain reexpands. This transition is absent with purely repulsive crowders, which produce only monotonic compaction.

This mechanism differs from the depletion-driven collapse described by Kang \emph{et al.}~\cite{Kang2015PRL} in several respects. In depletion, crowders are excluded from the polymer interior, compaction arises from osmotic pressure, and the chain retains SAW statistics at all densities. In bridging, crowders accumulate within the polymer domain, the transition is controlled by connectivity rather than excluded volume, and SAW universality is broken. As shown in Table~\ref{tbl:dimless_parameter_x}, bridging drives collapse at $x < 1$, a regime where the Kang \emph{et al.} framework predicts negligible compaction; the two mechanisms thus operate in distinct regions of parameter space. The analogy with the ``gluonic'' bridging mechanism identified in polyelectrolyte reentrant condensation~\cite{Yong2024Biomacromolecules} is worth noting: in that system, shared multivalent ions form physical crosslinks between ionic monomers, and saturation of those crosslinks drives re-entry. Our neutral crowder system realizes the same principle geometrically rather than electrostatically, with saturation arising from the finite number of monomer binding sites rather than from charge inversion.

The reentrant phenomenology observed here mirrors that seen across several seemingly disparate systems: ionic bridging and charge inversion in polyelectrolyte condensation~\cite{Raspaud1998BJ,Olvera1995JCP,Zhang2008PRL}, stoichiometric aggregation and redissolution in protein-ligand systems~\cite{Zhang2008PRL,Dwivedi2022MicrobeNotes}, and bridging flocculation followed by steric stabilization in colloid-polymer mixtures~\cite{ScheutjensFleer1979,Tripathy2017AdvColloid}. In each case, reentrance arises because multivalent interactions are most effective within a finite concentration window. Our results show that neither electrostatics nor solvent competition is required for this topology to emerge; saturable geometric bridging alone is sufficient.

The extension to charged polymers reveals that electrostatics amplifies rather than suppresses reentrance. For both $Z = 1$ and $Z = 3$, bridging crowders drive collapse at low $\phi_c$ despite intrachain Coulomb repulsion. Upon saturation of bridging at high $\phi_c$, the backbone charges, which have been held in proximity during the collapsed phase, experience restored mutual repulsion that drives the chain well beyond its original size ($R_g(\phi_c)/R_g(0) \approx 2.0$ for $Z = 1$, $\lambda = 3.6$). In this sense, the collapsed bridging phase acts as a conformational capacitor: electrostatic repulsion energy is stored during collapse and released upon saturation. The amplitude of this effect depends on the degree of counterion displacement: for $Z = 1$, bridging crowders fully displace counterions from the chain ($n_\mathrm{ci} \to 0$), leaving the backbone charges unscreened during reexpansion; for $Z = 3$, counterion displacement is incomplete due to stronger condensation at higher valency, and the retained counterion cloud partially screens the backbone charge, producing a more moderate normalized overshoot. The chain statistics reflect this amplification: at high $\phi_c$, $P(t)$ shifts to the right of the SAW master curve, defining a super-SAW regime with no neutral-chain analog.

Taken together, these results suggest that reentrant polymer behavior requires three ingredients: (i)~multivalent interactions capable of generating effective intrachain attraction; (ii)~a finite binding capacity that limits connectivity at high concentration; and (iii)~crowding levels sufficient to alter the balance between connectivity and steric competition. When these conditions are met, reentrance appears regardless of whether the driving interactions are electrostatic or purely steric. Purely repulsive crowders, which lack binding saturation, produce only monotonic responses~\cite{Kang2015PRL}. These conditions are satisfied not only in our minimal model but also in biological systems. For example, architectural chromatin-binding proteins such as condensins act as multivalent bridgers that compact chromosomes during mitosis, with bridging providing crosslinking connectivity that looping alone cannot achieve~\cite{Forte2024eLife}. The present framework suggests that the concentration dependence of such bridging proteins could generate reentrant chromosome compaction as bridging sites saturate at high protein density.

Future work may explore how polymer stiffness, polydisperse crowders, dynamic binding kinetics, and many-chain systems modify this picture. Extensions to intrinsically disordered proteins are particularly relevant: experiments have revealed non-monotonic conformational responses in crowded environments that depend on the sign and range of protein-crowder interactions~\cite{Balu2022SciAdv,Holehouse2024NatRevMolCellBiol}, and the present framework offers a natural mechanism for such behavior when crowder-residue interactions are attractive. Extensions to chromatin-like systems, where bridging proteins and electrostatics coexist and compete, may reveal new regimes of concentration-controlled structural transitions relevant to gene regulation. More broadly, the results suggest that reentrant conformational control via crowder concentration alone may be useful in designing responsive polymeric and biomolecular materials, where collapse and re-expansion can be triggered by a single component at fixed interaction strength simply by tuning its concentration.

\bibliography{neutral-reentrant}

@article{milin2018reentrant,
  title={Reentrant phase transitions and non-equilibrium dynamics in membraneless organelles},
  author={Milin, Anthony N and Deniz, Ashok A},
  journal={Biochemistry},
  volume={57},
  number={17},
  pages={2470--2477},
  year={2018},
  publisher={ACS Publications}
}

@article{banerjee2017reentrant,
  title={Reentrant phase transition drives dynamic substructure formation in ribonucleoprotein droplets},
  author={Banerjee, Priya R and Milin, Anthony N and Moosa, Mahdi Muhammad and Onuchic, Paulo L and Deniz, Ashok A},
  journal={Angewandte Chemie International Edition},
  volume={56},
  number={38},
  pages={11354--11359},
  year={2017},
  publisher={Wiley Online Library}
}

@article{feng2015re,
  title={Re-entrant solidification in polymer--colloid mixtures as a consequence of competing entropic and enthalpic attractions},
  author={Feng, Lang and Laderman, Bezia and Sacanna, Stefano and Chaikin, Paul},
  journal={Nature materials},
  volume={14},
  number={1},
  pages={61--65},
  year={2015},
  publisher={Nature Publishing Group UK London}
}

@article{truzzolillo2018overcharging,
  title={Overcharging and reentrant condensation of thermoresponsive ionic microgels},
  author={Truzzolillo, Domenico and Sennato, Simona and Sarti, Stefano and Casciardi, Stefano and Bazzoni, Chiara and Bordi, Federico},
  journal={Soft Matter},
  volume={14},
  number={20},
  pages={4110--4125},
  year={2018},
  publisher={Royal Society of Chemistry}
}

@article{marzi2015depletion,
  title={Depletion, melting and reentrant solidification in mixtures of soft and hard colloids},
  author={Marzi, Daniela and Capone, Barbara and Marakis, John and Merola, Maria Consiglia and Truzzolillo, Domenico and Cipelletti, Luca and Moingeon, Firmin and Gauthier, Mario and Vlassopoulos, Dimitris and Likos, Christos N and others},
  journal={Soft Matter},
  volume={11},
  number={42},
  pages={8296--8312},
  year={2015},
  publisher={Royal Society of Chemistry}
}

@article{Meng2017NPJ,
  author  = {Meng, Guangnan and Paulose, Jayson and Nelson, David R. and Manoharan, Vinothan N.},
  title   = {Elastic instability of a crystal growing on a curved surface},
  journal = {Science},
  volume  = {343},
  pages   = {634--637},
  year    = {2014},
 }

@article{LAMMPS,
  author = "A. P. Thompson and H. M. Aktulga and R. Berger and 
     D. S. Bolintineanu and W. M. Brown and P. S. Crozier and
     P. J. in 't Veld and A. Kohlmeyer and S. G. Moore and T. D. Nguyen and
     R. Shan and M. J. Stevens and J. Tranchida and C. Trott and S. J. Plimpton",
  title = "{LAMMPS} - a flexible simulation tool for
     particle-based materials modeling at the 
     atomic, meso, and continuum scales",
  journal = "Comp. Phys. Comm.",
  volume =  "271",
  pages =   "108171",
  year =    "2022",
  doi = "10.1016/j.cpc.2021.108171"
}

@article{Holehouse2024NatRevMolCellBiol,
  author  = {Holehouse, Alex S. and Kragelund, Birthe B.},
  title   = {The molecular basis for cellular function of intrinsically 
             disordered protein regions},
  journal = {Nature Reviews Molecular Cell Biology},
  volume  = {25},
  pages   = {187--211},
  year    = {2024}
}

@book{DeGennes1979,
  author    = {de Gennes, Pierre-Gilles},
  title     = {Scaling Concepts in Polymer Physics},
  publisher = {Cornell University Press},
  address   = {Ithaca},
  year      = {1979}
}

@article{Lhuillier1988JPhys,
  author  = {Lhuillier, D.},
  title   = {A simple model for polymeric fractals in a good solvent 
             and an improved version of the {F}lory approximation},
  journal = {Journal de Physique},
  volume  = {49},
  pages   = {705--710},
  year    = {1988}
}

@article{Yong2024Biomacromolecules,
  author  = {Yong, Hao},
  title   = {Reentrant Condensation of Polyelectrolytes by Multivalent 
             Ions: {G}luonic Effect and Charge Inversion},
  journal = {Biomacromolecules},
  volume  = {25},
  pages   = {7361--7372},
  year    = {2024}
}

@article{Manning1969JCP,
  author  = {Manning, Gerald S.},
  title   = {Limiting Laws and Counterion Condensation in Polyelectrolyte 
             Solutions I. Colligative Properties},
  journal = {The Journal of Chemical Physics},
  volume  = {51},
  pages   = {924--933},
  year    = {1969}
}

@article{srivastava2016polyelectrolyte,
  title={Polyelectrolyte complexation},
  author={Srivastava, Samanvaya and Tirrell, Matthew V},
  journal={Advances in chemical physics},
  volume={161},
  pages={499--544},
  year={2016},
  publisher={Wiley Online Library}
}

@book{GroKho1994,
  author    = {Grosberg, Alexander Yu. and Khokhlov, Alexei R.},
  title     = {Statistical Physics of Macromolecules},
  publisher = {AIP Press},
  address   = {New York},
  year      = {1994}
}

@book{RubinsteinColby2003,
  author    = {Rubinstein, Michael and Colby, Ralph H.},
  title     = {Polymer Physics},
  publisher = {Oxford University Press},
  address   = {Oxford},
  year      = {2003}
}

@article{grassmann2024computational,
  title={Computational approaches to predict protein--protein interactions in crowded cellular environments},
  author={Grassmann, Greta and Miotto, Mattia and Desantis, Fausta and Di Rienzo, Lorenzo and Tartaglia, Gian Gaetano and Pastore, Annalisa and Ruocco, Giancarlo and Monti, Michele and Milanetti, Edoardo},
  journal={Chemical Reviews},
  volume={124},
  number={7},
  pages={3932--3977},
  year={2024},
  publisher={ACS Publications}
}

@article{alfano2024molecular,
  title={Molecular crowding: the history and development of a scientific paradigm},
  author={Alfano, Caterina and Fichou, Yann and Huber, Klaus and Weiss, Matthias and Spruijt, Evan and Ebbinghaus, Simon and De Luca, Giuseppe and Morando, Maria Agnese and Vetri, Valeria and Temussi, Piero Andrea and others},
  journal={Chemical Reviews},
  volume={124},
  number={6},
  pages={3186--3219},
  year={2024},
  publisher={ACS Publications}
}

@article{garg2025polymer,
  title={Polymer Conformations with Attractive Bridging Crowder Interactions: Role of Crowder Size},
  author={Garg, Hitesh and Vemparala, Satyavani},
  journal={Macromolecules},
  volume={58},
  number={20},
  pages={11006--11016},
  year={2025},
  publisher={ACS Publications}
}

@article{rivas2016macromolecular,
  title={Macromolecular crowding in vitro, in vivo, and in between},
  author={Rivas, Germ{\'a}n and Minton, Allen P},
  journal={Trends in biochemical sciences},
  volume={41},
  number={11},
  pages={970--981},
  year={2016},
  publisher={Elsevier}
}

@article{Forte2024eLife,
  author  = {Forte, Giada and others},
  title   = {Bridging condensins mediate compaction of mitotic chromosomes},
  journal = {eLife},
  volume  = {13},
  pages   = {e79917},
  year    = {2024},
  doi     = {10.7554/eLife.79917}
}

@article{Balu2022SciAdv,
  author  = {Balu, Rajkamal and others},
  title   = {Crowder-directed interactions and conformational dynamics in 
             multistimuli-responsive intrinsically disordered protein},
  journal = {Science Advances},
  volume  = {8},
  pages   = {eabq2202},
  year    = {2022},
  doi     = {10.1126/sciadv.abq2202}
}

@article{Mukherji2014NatComm,
  author  = {Mukherji, Debashish and Marques, Carlos M. and Kremer, Kurt},
  title   = {Polymer collapse in miscible good solvents is a generic phenomenon 
             driven by preferential adsorption},
  journal = {Nature Communications},
  volume  = {5},
  pages   = {4882},
  year    = {2014},
  doi     = {10.1038/ncomms5882}
}

@article{Scherzinger2022SoftMatter,
  author  = {Scherzinger, Christine and Ballauff, Matthias and Richtering, Walter 
             and Papadakis, Christine M.},
  title   = {Cononsolvency of thermoresponsive polymers: where we are now 
             and where we are going},
  journal = {Soft Matter},
  volume  = {18},
  pages   = {3553},
  year    = {2022},
  doi     = {10.1039/D2SM00146B}
}

@article{Kang2015PRL,
  author  = {Kang, Hongsuk and Pincus, Philip A. and Hyeon, Changbong 
             and Thirumalai, D.},
  title   = {Effects of Macromolecular Crowding on the Collapse of Biopolymers},
  journal = {Physical Review Letters},
  volume  = {114},
  pages   = {068303},
  year    = {2015},
  doi     = {10.1103/PhysRevLett.114.068303}
}

@article{HuangCheng2021JPolSci,
  author  = {Huang, Yishan and Cheng, Shengfeng},
  title   = {Reentrant Transition of Polymer Collapse in an Athermal 
             Solvent with Attractive Polymer--Solvent Interactions},
  journal = {Journal of Polymer Science},
  volume  = {59},
  pages   = {2819},
  year    = {2021},
  doi     = {10.1002/pol.20210486}
}

@article{Garg2023JCP,
  author  = {Garg, Hitesh and Rajesh, R. and Vemparala, Satyavani},
  title   = {The conformational phase diagram of neutral polymers in the 
             presence of attractive crowders},
  journal = {The Journal of Chemical Physics},
  volume  = {158},
  pages   = {114903},
  year    = {2023},
  doi     = {10.1063/5.0140721}
}

@article{Tripathi2023JCP,
  author  = {Tripathi, Kamal and Garg, Hitesh and Rajesh, R. and Vemparala, Satyavani},
  title   = {The conformational phase diagram of charged polymers in the 
             presence of attractive bridging crowders},
  journal = {The Journal of Chemical Physics},
  volume  = {159},
  pages   = {204903},
  year    = {2023},
  doi     = {10.1063/5.0172696}
}

@article{Nayar2020PCCP,
  author  = {Nayar, Divya},
  title   = {Small crowder interactions can drive hydrophobic polymer collapse 
             as well as unfolding},
  journal = {Physical Chemistry Chemical Physics},
  volume  = {22},
  pages   = {18091},
  year    = {2020},
  doi     = {10.1039/d0cp02402c}
}

@article{Nayar2023JPCB,
  author  = {Nayar, Divya},
  title   = {Molecular Crowders Can Induce Collapse in Hydrophilic Polymers 
             via Soft Attractive Interactions},
  journal = {The Journal of Physical Chemistry B},
  volume  = {127},
  pages   = {6265},
  year    = {2023},
  doi     = {10.1021/acs.jpcb.3c01319}
}

@article{Zhang2001PRL,
  title={Reentrant coil–globule–coil transition of poly(N-isopropylacrylamide) in water–methanol mixtures},
  author={Zhang, G. and Wu, C.},
  journal={Physical Review Letters},
  volume={86},
  pages={822--825},
  year={2001}
}

@article{Okay2021Gels,
  title={Reentrant volume phase transitions in responsive hydrogels},
  author={Okay, Oguz},
  journal={Gels},
  volume={7},
  pages={98},
  year={2021}
}

@article{oh2013molecular,
  title={Molecular thermodynamic analysis for reentrant and reentrant-convex type swelling behaviors of thermo-sensitive hydrogels in mixed solvents},
  author={Oh, Suk Yung and Bae, Young Chan},
  journal={Polymer},
  volume={54},
  number={9},
  pages={2308--2314},
  year={2013},
  publisher={Elsevier}
}

@article{Raspaud1998BJ,
  title={Precipitation of DNA by polyamines: a polyelectrolyte behavior},
  author={Raspaud, E. and Olvera de la Cruz, M. and Sikorav, J. L. and Livolant, F.},
  journal={Biophysical Journal},
  volume={74},
  pages={381--393},
  year={1998}
}

@article{Olvera1995JCP,
  title={Complexation of polylectrolytes with multivalent ions: theory and experiment},
  author={Olvera de la Cruz, M. and Belloni, L. and Delsanti, M. and Dalbiez, J. P. and Spalla, O. and Drifford, M.},
  journal={Journal of Chemical Physics},
  volume={103},
  pages={5781--5791},
  year={1995}
}

@article{Liu2020JCIS,
  title={Re-entrant swelling and redissolution of polyelectrolytes in concentrated salt solutions},
  author={Liu, J. and others},
  journal={Journal of Colloid and Interface Science},
  volume={579},
  pages={369--379},
  year={2020}
}

@article{Robertson2023PCCP,
  title={Underscreening-driven reentrant swelling of polyelectrolyte brushes at high salt},
  author={Robertson, M. B. and others},
  journal={Physical Chemistry Chemical Physics},
  volume={25},
  pages={24770--24782},
  year={2023}
}

@article{Zhang2008PRL,
  title={Reentrant condensation of proteins with multivalent cations},
  author={Zhang, F. and others},
  journal={Physical Review Letters},
  volume={101},
  pages={148101},
  year={2008}
}

@misc{Dwivedi2022MicrobeNotes,
  title={Antigen–antibody precipitation reactions},
  author={Dwivedi, A.},
  howpublished={\url{https://microbenotes.com/}},
  year={2022}
}

@article{ScheutjensFleer1979,
  title={Statistical theory of the adsorption of interacting chain molecules. 1. Partition function, segment density distribution, and adsorption isotherms},
  author={Scheutjens, J. M. H. M. and Fleer, G. J.},
  journal={Journal of Physical Chemistry},
  volume={83},
  pages={1619--1635},
  year={1979}
}

@article{Tripathy2017AdvColloid,
  title={Polymer bridging flocculation: principles and applications},
  author={Tripathy, T. and De, B. R.},
  journal={Advances in Colloid and Interface Science},
  volume={247},
  pages={62--81},
  year={2017}
}

@article{mukherji2017depleted,
  title={Depleted depletion drives polymer swelling in poor solvent mixtures},
  author={Mukherji, Debashish and Marques, Carlos M and Stuehn, Torsten and Kremer, Kurt},
  journal={Nature communications},
  volume={8},
  number={1},
  pages={1--7},
  year={2017},
  publisher={Nature Publishing Group}
}

@incollection{lekkerkerker2011depletion,
  title={Depletion interaction},
  author={Lekkerkerker, Henk NW and Tuinier, Remco},
  booktitle={Colloids and the depletion interaction},
  pages={57--108},
  year={2011},
  publisher={Springer}
}

@article{ellis2003join,
  title={Join the crowd},
  author={Ellis, R John and Minton, Allen P},
  journal={Nature},
  volume={425},
  number={6953},
  pages={27--28},
  year={2003},
  publisher={Nature Publishing Group}
}

@article{zimmerman1993macromolecular,
  title={Macromolecular crowding: biochemical, biophysical, and physiological consequences},
  author={Zimmerman, Steven B and Minton, Allen P},
  journal={Annual review of biophysics and biomolecular structure},
  volume={22},
  number={1},
  pages={27--65},
  year={1993},
  publisher={Annual Reviews 4139 El Camino Way, PO Box 10139, Palo Alto, CA 94303-0139, USA}
}

@article{zhou2008macromolecular,
  title={Macromolecular crowding and confinement: biochemical, biophysical, and potential physiological consequences},
  author={Zhou, Huan-Xiang and Rivas, Germ{\'a}n and Minton, Allen P},
  journal={Annu. Rev. Biophys.},
  volume={37},
  pages={375--397},
  year={2008},
  publisher={Annual Reviews}
}
\end{document}